
%
%
%
%
\input harvmac
\input tables
\ifx\epsfbox\notincluded\message{(NO epsf.tex, FIGURES WILL BE IGNORED)}
\def\insertfig#1#2{}
\else\message{(FIGURES WILL BE INCLUDED)}
\def\insertfig#1#2{
\midinsert\centerline{\epsffile{#2}}\centerline{{#1}}\endinsert}
\fi
%
%
%
%
\ifx\answ\bigans
\else
\output={
  \almostshipout{\leftline{\vbox{\pagebody\makefootline}}}\advancepageno
}
\fi
%
%
%

%
%

%
%
\def\UCSD#1#2{\noindent#1\hfill #2%
\supereject\global\hsize=\hsbody%
\footline={\hss\tenrm\folio\hss}}
%
%
\def\abstract#1{\centerline{\bf Abstract}\nobreak\medskip\nobreak\par #1}
%
%
%
%
\edef\tfontsize{ scaled\magstep3}
 \tfontsize  \tfontsize
 \tfontsize \font\titlei=cmmi10 \tfontsize
\font\titleis=cmmi7 \tfontsize \font\titleiss=cmmi5 \tfontsize
\font\titlesy=cmsy10 \tfontsize \font\titlesys=cmsy7 \tfontsize
\font\titlesyss=cmsy5 \tfontsize  \tfontsize
\skewchar\titlei='177 \skewchar\titleis='177 \skewchar\titleiss='177
\skewchar\titlesy='60 \skewchar\titlesys='60 \skewchar\titlesyss='60
%
%
%
%
%
\def\inv{^{\raise.15ex\hbox{${\scriptscriptstyle -}$}\kern-.05em 1}}
\def\lbar{{\lower.35ex\hbox{$\mathchar'26$}\mkern-10mu\lambda}} 

%
%
%
%
\def\slash#1{\rlap{$#1$}/} 
\def\dsl{\,\raise.15ex\hbox{/}\mkern-13.5mu D} 
\def\delsl{\raise.15ex\hbox{/}\kern-.57em\partial}
\def\Ksl{\hbox{/\kern-.6000em\rm K}}
\def\Asl{\hbox{/\kern-.6500em \rm A}}
\def\Dsl{\hbox{/\kern-.6000em\rm D}} 
\def\Qsl{\hbox{/\kern-.6000em\rm Q}}
\def\gradsl{\hbox{/\kern-.6500em$\nabla$}}
%
%
\def\lspace{\ifx\answ\bigans{}\else\qquad\fi}
\def\lbspace{\ifx\answ\bigans{}\else\hskip-.2in\fi} 
%
%
\def\boxeqn#1{\vcenter{\vbox{\hrule\hbox{\vrule\kern3pt\vbox{\kern3pt
        \hbox{${\displaystyle #1}$}\kern3pt}\kern3pt\vrule}\hrule}}}
%
%
\def\mbox#1#2{\vcenter{\hrule \hbox{\vrule height#2in
\kern#1in \vrule} \hrule}}
%
%
%
%
   
   \def\CH{{\cal H}}
   \def\CL{{\cal L}}

%
%
%
%
%

%

\def\bar#1{\overline{#1}}

\def\bra#1{\left\langle #1\right|}
\def\ket#1{\left| #1\right\rangle}

\def\darr#1{\raise1.5ex\hbox{$\leftrightarrow$}\mkern-16.5mu #1}

%
%
\def\frac#1#2{{\textstyle{#1\over #2}}} 
%
%
%
%

\def\Tr{\mathop{\rm Tr}}

%
%
%
%

%
%
\def\ltap{\ \raise.3ex\hbox{$<$\kern-.75em\lower1ex\hbox{$\sim$}}\ }
\def\gtap{\ \raise.3ex\hbox{$>$\kern-.75em\lower1ex\hbox{$\sim$}}\ }
\def\gl{\ \raise.5ex\hbox{$>$}\kern-.8em\lower.5ex\hbox{$<$}\ }
\def\roughly#1{\raise.3ex\hbox{$#1$\kern-.75em\lower1ex\hbox{$\sim$}}}
%
%
        
\def\eg{\hbox{\it e.g.}}        
\def\etal{\hbox{\it et al.}}

\def\np#1#2#3{Nucl. Phys. B{#1} (#2) #3}
\def\pl#1#2#3{Phys. Lett. {#1}B (#2) #3}

\def\physrev#1#2#3{Phys. Rev. {#1} (#2) #3}

\relax
\noblackbox
\centerline{{\titlefont Exotic $QQ\bar q \bar q$ States in QCD}}
\bigskip
\centerline{Aneesh V. Manohar}
\centerline{{\sl CERN TH-Division, CH-1211 Geneva 23,
Switzerland}\footnote{*}{On leave from the University of California at
San Diego.}}
\medskip
\centerline{Mark B. Wise}
\centerline{{\sl California Institute of Technology, Pasadena, CA 91125}}
\vfill
\abstract{We show that QCD contains stable four-quark $QQ\bar q \bar q$
hadronic states in the limit where the heavy quark mass goes to infinity.
(Here $Q$ denotes a heavy quark, $\bar q$ a light antiquark and the stability
refers only to the strong interactions.) The long range binding
potential is due to one pion exchange between ground state $Q\bar q$
mesons, and is computed using chiral perturbation theory. For the $Q=b$,
this long range potential may be sufficiently attractive to produce a
weakly bound two meson state.}
\vfill
\UCSD{\vbox{\hbox{CERN-TH.6744/92}\hbox{CALT-68-1869}}}{December 1992}

\newsec{Introduction}

QCD with three light quark flavors contains  mesons (with $\bar qq$
flavor quantum numbers) and baryons (with $qqq$ flavor quantum numbers) in its
spectrum.  In addition there are ``nuclei'' which are weakly bound
states of the baryons.  At the present time there is no evidence for
other types of states that are stable with respect to the strong interactions.
It was originally suggested by Jaffe \ref\jaffe{R.L. Jaffe,
\physrev{D15}{1977}{267}, \physrev{D15}{1977}{281}\semi
J. Weinstein and N. Isgur, \physrev{D41}{1990}{2236}}\
that there should be hadronic resonances with $qq \bar q
\bar q$ flavor quantum numbers and there is evidence that some of the
observed hadronic resonances should be interpreted in this way.  In this
paper we study the possibility of {\it stable} exotic $QQ \bar q \bar q$
hadrons, where $Q$ is a heavy quark (i.e., $m_Q \gg \Lambda_{{\rm
QCD}}$) and the stability refers only to the strong interactions.  It is
easy to see that such states exist in the $m_Q \rightarrow \infty$
limit.  For very heavy quarks, the quark pair $QQ$ can form a small
color antitriplet object of size $\left(\alpha_s\left(m_Q\right)
m_Q\right)^{-1}$ with a binding energy of order
$\alpha_s^2\left(m_Q\right) m_Q$.   The two heavy quarks act as an
almost point-like  heavy color antitriplet source with a mass of about
$2m_Q$ for the two light antiquarks in the $QQ\bar q\bar q$ hadron. This
results in bound states that have the light degrees of freedom in
configurations similar to those in the observed  $\bar\Lambda_b$ and
$\bar\Lambda_c$ states,  with the $QQ$ pair playing the role of the
heavy antiquark. In the heavy quark limit, the binding energy of the
$QQ$ pair tends to infinity, whereas the energy of the light degrees of
freedom is of order $\Lambda_{\rm QCD}$, so that the $QQ\bar q\bar q$ state
has a lower energy than two separate $Q\bar q$ mesons
and is stable with respect to the strong interactions.

This argument for the existence of exotic $QQ \bar q \bar q$ states in
the spectrum of QCD is based on the short range color Coulomb attraction
in the channel $3 \otimes 3 \rightarrow \bar 3$.  For the case where $Q$
is the top quark, this description is likely to be quantitatively
correct and establishes the existence of exotic $tt\bar q  \bar q$
states that are stable with respect to the strong interactions. The
charm and bottom quarks are not heavy enough for their short range color
Coulomb attraction to play an important role in the formation of a
$QQ\bar q \bar q$ state, since $\alpha_s^2(m_Q)\,m_Q$ is not large
compared with $\Lambda_{\rm QCD}$. If $QQ\bar q \bar q$ states exist for
$Q = c$ or $b$, they may be weakly bound two meson systems and the
formation of such a bound state depends on the potential between the
lowest lying $Q\bar q$ mesons.  At long distances this potential is
determined by one pion exchange and is calculable in chiral perturbation
theory.  For the remainder of this paper we examine the picture of $QQ
\bar q \bar q$ hadrons as two weakly bound $Q\bar q$ mesons, and apply
it to the $c$ and $b$ quark systems.

Section 2 contains a derivation of the long range potential using chiral
perturbation theory.  In Section 3 the eigenstates of the potential
operator are classified and $\Lambda_{{\rm QCD}}/m_Q$ corrections to the
Hamiltonian  are discussed.  In Section 4 we present a variational
calculation that suggests there might be a weakly bound state involving
$B$ and $B^*$ mesons. Some concluding remarks are also given.

\newsec {The Long Range Potential Between Heavy Mesons}

In the limit $m_Q \rightarrow \infty$, the angular momentum of the light
degrees of freedom, $s_\ell$, is a good quantum number.  Mesons
containing a single heavy quark come in degenerate doublets with total
spins $s_\pm = s_\ell \pm 1/2$.  The ground state multiplets with $Q\bar
q_a$ flavor quantum numbers ($q_1 = u, q_2 = d, q_3 = s$) have $s_\ell =
1/2$ and negative parity, giving a doublet of pseudoscalar and vector
mesons that we denote by $P_a^{(Q)}$ and $P_a^{*(Q)}$ respectively.  For
$Q = c$ these  are the ($D^0, D^+, D_s$) and ($D^{*0}, D^{*+}, D_s^*$)
mesons and for $Q = b$ they are the ($B^-, \bar B^0, B_s$) and ($B^{*-},
\bar B^{*0}, B_s^*$) mesons.

The interactions of these heavy mesons with the $\pi, K$ and $\eta$ is
determined by the chiral and heavy quark symmetries of the strong
interactions.  The effective Lagrangian that describes the low momentum
strong interactions of the pseudo-Goldstone bosons with the $P_a^{(Q)}$
and $P_a^{*(Q)}$ mesons is \ref\hqlagref{M.B. Wise,
\physrev{D45}{1992}{2188}\semi G. Burdman and J. Donoghue,
\pl{280}{1992}{287}\semi T.M. Yan, \etal, \physrev{D46}{1992}{1148}}
\eqn\hqlag{\eqalign{
{\cal L} &= - i \Tr \bar H^{(Q)} v_\mu \partial^\mu H^{(Q)} + {i\over 2}
\Tr \bar H^{(Q)} H^{(Q)} v^\mu [\xi^{\dagger} \partial_\mu \xi + \xi
\partial_\mu \xi^{\dagger}]\cr
&   + {ig\over 2} \Tr \bar H^{(Q)} H^{(Q)} \gamma_\nu \gamma_5
[\xi^{\dagger} \partial^\nu \xi - \xi \partial^\nu \xi^{\dagger}]\cr
&  + \lambda_1 \Tr \bar H^{(Q)} H^{(Q)} [\xi m_q \xi +
\xi^{\dagger} m_q \xi^{\dagger}] + \lambda'_1 \Tr \bar H^{(Q)}
H^{(Q)} \Tr [m_q \Sigma + m_q \Sigma^{\dagger}]\cr
&+ {\lambda_2\over m_Q} \Tr \bar H^{(Q)} \sigma^{\mu\nu}
H^{(Q)} \sigma_{\mu\nu} + ... \,\,,
}}
where the ellipsis denotes terms with more derivatives, more factors of
the light quark mass matrix
\eqn\qmass{
        m_q = \pmatrix{m_u & 0 & 0\cr
        0 & m_d & 0\cr
        0 & 0 & m_s\cr} \,\, ,
}
or more factors of $1/m_Q$ associated with violation of heavy quark spin
symmetry. The traces in eq.~\hqlag\ are over flavor and spinor indices.
The pseudoscalar and vector heavy meson fields $P_a^{(Q)}$,
$P_{a\mu}^{*(Q)}$  are combined to form the $4\times 4$ Lorentz bispinor matrix
\eqn\hfield{
H_a^{(Q)} = {\left(1 + \slash v\right)\over 2}
[P_{a\mu}^{*(Q)} \gamma^\mu - P_a^{(Q)} \gamma_5] \,\, .
}
The field $H^{(Q)}$ destroys $P^{(Q)}$ and $P^{*(Q)}$ mesons with
four-velocity $v^\mu$.  The subscript $v$ on $H^{(Q)}$, $P^{(Q)}$ and
$P^{*(Q)}$ has been omitted to simplify the notation.  The conjugate
``barred'' field is defined by
\eqn\hbfield{
\bar H^{(Q)a} = \gamma^0 H_a^{(Q)\dag} \gamma^0 \,\, .
}

The pseudo-Goldstone bosons appear in the Lagrangian through
\eqn\xidef{
\xi = \exp \left({iM\over f}\right) \,\, ,
}
where
\eqn\Mdef{
M = \left[\matrix{{1\over \sqrt{2}} \pi^0 + {1\over \sqrt{6}}
\eta & \pi^+ & K^+\cr
\pi^- & - {1\over \sqrt{2}} \pi^0 + {1\over \sqrt{6}} \eta & K^0\cr
K^- & \bar K^0 & - {2\over\sqrt{6}} \eta\cr}\right] \,\, ,
}
and
\eqn\sigmadef{
\Sigma = \xi^2 \,\, .
}
In eq.~\xidef, $f$ is the pion decay constant, $f \simeq$ 132 MeV.

Under $SU(3)_L \times SU(3)_R$ chiral symmetry transformations
\eqn\sigtrans{
\Sigma \rightarrow L \Sigma R^{\dag} \,\, ,
}
where $L\in SU(3)_L$ and $R\in SU(3)_R$.  The transformation
law of $\xi$ is then
\eqn\xitrans{
\xi \rightarrow L \xi U^{\dag} = U \xi R^{\dag} \,\, ,
}
where $U$ is a complicated function of $L,R$ and the mesons $M$. In
general $U$ depends on space-time, but for transformations in the
unbroken $SU(3)_V$ subgroup $V = L = R$, $\xi \rightarrow V \xi
V^{\dag}$, and $U$ is the constant matrix $V$.  Under $SU(2)_v$ heavy
quark spin symmetry and $SU(3)_L \times SU(3)_R$ chiral symmetry, the
heavy meson fields transform as
\eqn\htrans{
 H^{(Q)} \rightarrow S\,H^{(Q)}
U^{\dag} \,\, ,
}
where $S\in SU(2)_v$ is the heavy quark spin transformation.

The long range potential between heavy mesons is determined by the one
pion exchange Feynman diagram in \fig\figone{The one-pion exchange contribution
to the meson-meson potential. The solid lines are either the $P^{(Q)}$
or $P^{*(Q)}$, and the dashed line is the pion.}, with the virtual pion
having momentum transfer $q^\mu = \left(0,\vec q\right)$.  At low
momentum, the coupling of the $P^{(Q)}$-$P^{*(Q)}$ and
$P^{*(Q)}$-$P^{*(Q)}$ to the Goldstone bosons is obtained by expanding
the Lagrangian eq.~\hqlag\ up to first order in the pion fields. The
only term which contributes to the one-pion coupling at low momentum
transfer is the term proportional to $g$ in eq.~\hqlag. In the $m_Q
\rightarrow \infty$ limit the $P^{(Q)}$ and $P^{*(Q)}$ are degenerate in
mass  and can be treated as a single ``$H$ particle.'' This gives the
$H$-pion interaction
\eqn\intlag{
\CL_{\rm int}=-{g\over f}
\Tr \bar H^{(Q)} H^{(Q)} \gamma_\nu \gamma_5\ \partial^\nu\pi.
}
The interaction can be rexpressed in terms of the spin of the light
degrees of freedom $\vec S_\ell$ and the isopsin $I^A$ of the $H$ field as
\eqn\newint{
\CL_{\rm int}={2\sqrt{2}g\over f}\ \left(\vec S_\ell\cdot
\vec \partial\pi^A\right)\ I^A.
}
The Fourier transform of the interaction potential obtained from eq.~\newint
\ between two $H$ particles is
\eqn\kpot{
V_\pi\left(\vec q\right) = - {8 g^2\over f^2}\ \vec I_1 \cdot \vec I_2\
{\left(\vec S_{\ell 1} \cdot \vec q\right)
\left(\vec S_{\ell 2} \cdot \vec q\right)\over \vec q^{\ 2}+
m_\pi^2} \,\, .
}
$\vec I_{1,2}$ denotes the isospin of heavy meson $1$ and $2$ and $\vec
S_{\ell 1,2}$ denotes the spin of the light degrees of freedom in heavy
meson $1$ and $2$.  In coordinate space, eq.~\kpot\ gives the potential
\eqn\pot{
V_\pi\left(\vec x\right) = 4 \vec I_1 \cdot \vec I_2
\Big[\left(\vec S_{\ell 1} \cdot \hat x
\ \vec S_{\ell 2} \cdot \hat x - {1\over 3} \ \vec S_{\ell 1} \cdot \vec
S_{\ell 2}\right)\ W_2 \left(r\right)
+ \vec S_{\ell 1} \cdot \vec S_{\ell 2}\ W_0 \left(r\right)\Big] \,\, ,
}
where
\eqn\wtwo{
W_2\left(r\right) = {g^2\over 2\pi f^2}\ e^{-m_{\pi}r} \left({3\over r^3} +
{3m_\pi \over r^2} + {m_\pi^2\over r}\right) \,\, ,
}
and
\eqn\wzero{
W_0 \left(r\right) = {g^2\over 2\pi f^2}\ e^{-m_{\pi}r}
\left({m_\pi^2\over
3 r}\right) \,\, .
}

The coupling $g$ determines the $D^* \rightarrow D\pi$ decay rate.  At
tree level
\eqn\width{
\Gamma \left(D^{*+} \rightarrow D^0 \pi^+\right) = {g^2\over 6\pi f^2}
|\vec p_\pi|^3 \,\, .
}
The decay width for $D^{*+} \rightarrow D^+ \pi^0$ is a factor of two
smaller (this follows from isospin invariance).  The experimental upper
limit on the $D^{*+}$ width of 131 keV \ref\expt{The ACCMOR collaboration,
S. Barlag \etal, \pl{278}{1992}{480}}\ combined with the measured
branching ratios for $D^{*+} \rightarrow D^+ \pi^0$ and $D^{*+}
\rightarrow D^0 \pi^+$ leads to the limit $g^2 \ltap 0.5$.  A
measurement of the branching ratio for $D^{*+} \rightarrow D^+ \gamma$
could also lead to valuable information on $g$
\ref\rosner{J. Amundson, \etal, CERN Preprint CERN-TH.6650/92 (1992)\semi
P. Cho and H. Georgi, Harvard University Preprint HUTP-92/A043 (1992)}.
The axial current obtained from the Lagrange density eq.~\hqlag\ is
\eqn\axial{
\bar q\ T^A\, \gamma_\nu \gamma_5\, q = - g\ \Tr \bar H\, H\,
\gamma_\nu \gamma_5\, T^A + ... \,\, ,
}
where the ellipsis denotes terms containing one or more Goldstone boson
fields and $T^A$ is a flavor $SU(3)$ generator.  Treating the quark fields
in eq.~\axial\ as constituent quarks and using the nonrelativistic
constituent quark model to estimate the $D^*$ matrix element of the
l.h.s. of eq.~\axial\ gives $g = 1$. (A similar estimate of the pion nucleon
coupling constant gives $g_A = 5/3$.)  In the chiral quark model there
is a constituent quark pion coupling \ref\georgi{A.V. Manohar and H. Georgi,
\np{234}{1984}{189}}.  Using the measured
pion-nucleon coupling to determine the constituent quark pion coupling
gives $g^2 \simeq 0.6$.  Thus, various constituent quark model
calculations lead to the expectation that $g$ is near unity.

Equation~\pot\ is the leading contribution to the long distance part of
the heavy meson interaction potential.  Loops and higher derivative
operators give contributions that are suppressed by factors of $(4 \pi
fr)^{-1}$.  It seems reasonable that eqs.~\pot--\wzero\ dominate the
potential at distances greater than $r_{\rm min} \equiv (1/2 m_\pi)
\simeq 0.7$~fm. In the nuclear potential the corrections to one pion
exchange are important even at this distance
\ref\nuc{S.O. Backman, G.E. Brown and J.A. Niskanen,
Phys. Rep. 124 (1985) 1}. We believe this is (at
least partly) due to integrating out the $\Delta$ resonance which has a
large coupling to $N\pi$.  In the heavy meson case both the $P^{(Q)}$
and $P^{*(Q)}$ mesons are kept in the lagrangian~\hqlag. The lightest
heavy mesons that are integrated out of the theory do not couple very
strongly to $P^{(Q)}\pi$ and $P^{*(Q)}\pi$ since in the constituent
quark model they correspond to $P$-wave orbital excitations. At
$r=(2m_\pi)^{-1}$, $W_2 \left(r_{\rm min}\right) = 518 g^2$~MeV and $W_0
\left(r_{\rm min}\right) =  9 g^2$~MeV, where we have used the neutral
pion mass in computing the numerical values.

The eigenvalues of the potential operator eq.~\pot\ are easily
determined.  The position space part of the state vector can be taken to
be $|\hat z\rangle$, corresponding to the spatial wavefunction $\delta^3
\left(\vec x - r \hat z\right)$, since the potential is rotationally
invariant. Then eigenstates of $V_\pi\left(\vec x\right)$ are  $|I\,
I^3\rangle| \hat z\rangle |K\,k\rangle$ where
\eqn\kdef{
\vec K = \vec S_{\ell 1} + \vec S_{\ell 2} \,\, ,
}
is the total spin of the light degrees of freedom (the spin quantization
axis is also taken to be the $z$-axis) and
\eqn\itotal{
\vec I = \vec I_1 + \vec I_2 \,\, ,
}
is the total isospin.  Acting on these eigenstates the potential energy
eq.~\pot\ can be rewritten as
\eqn\newpot{\eqalign{
V_\pi &= \left(I^2-I_1^2-I_2^2\right) \Biggl[\left(2 S_{\ell 1}^z
\ S_{\ell 2}^z\ W_2\left(r\right) - {1\over 3} \ \vec S_{\ell 1} \cdot \vec
S_{\ell 2}\right)\ W_2 \left(r\right)\cr
&\qquad+\left(K^2-S_{\ell 1}^2-S_{\ell 2}^2\right)
\ \left( W_0 \left(r\right)-
{1\over3}W_2\left(r\right)\right)\Biggr] \,\, .
}}
$S_{\ell 1}^z S_{\ell 2}^z
$ is $1/4$ for states with $k=\pm1$, and is
$-1/4$ for states with $k=0$, so we obtain the eigenvalues of the
potential for the states  as given in Table~1,
\midinsert
\centerline{{\bf TABLE l}}
\nobreak\bigskip\nobreak
\begintable ~~$I$~~|~~$K$~~|~~$k$~~|~~$V_\pi$~~\crthick
0|0|0|$9W_0/4$\cr
0|1|0|$3X_0$\cr
0|1|$\pm 1$|~~$-3X_1$~~\cr
1|0|0|$-3W_0/4$\cr
1|1|0|$-X_0$\cr
1|1|$\pm 1$|$X_1$\endtable
\endinsert
\noindent where $X_0$ and $X_1$ are defined by
\eqn\xzero{
X_0 = {W_2\over 3} - {W_0\over 4} \,\, ,
}
and
\eqn\xone{
X_1 = {W_2\over 6} + {W_0\over 4} \,\, .
}
Since $W_0$, $X_0$ and $X_1$ are positive, the attractive channels have
$\left(I, K, |k|\right) = \left(0,1,1\right), \left(1,0,0\right)$ and
$\left(1,1,0\right)$.  At $r_{\rm min}$  the potential energies for
these states are about $-266 g^2$~MeV, $-7g^2$~MeV and $-170 g^2$~MeV
respectively.

\newsec {Classification of Eigenstates}

In the previous section we found spatial $\otimes$ spin parts of the
eigenstates of the potential $V$ that were of the form $|\hat
z\rangle|K\,k\rangle$. By rotational invariance $\hat
R\left(g\right)\left[\ |\hat z\rangle|K\,k\rangle\ \right]$ is an
eigenstate of  $V$ with the same energy for any rotation $g$.  It is
convenient to combine these states into ones with a definite ``angular
momentum of the light degrees of freedom'' $\vec F$ using\foot{$F$ is
the total angular momentum minus the spin of the heavy quarks. It is not
the true angular momentum of the light degrees of freedom because it
contains the orbital angular momentum of the heavy quarks.}\
\ref\skyrme{A.V. Manohar,  \np{248}{1984}{19}}
\eqn\jstate{
|F\,f; K\,k\rangle = \sqrt{2F+1} \int_{SU(2)}
dg\ D_{fk}^{(F)*} (g)\ \hat
R (g)\left[\ | \hat z\rangle|K\,k\rangle\ \right]\  ,
}
where $D_{fk}^{(F)} (g)$ is the rotation matrix in representation $F$
and the measure is chosen so that
\eqn\meas{
\int_{SU(2)} dg = 1 .
}
Alternatively we can combine states with definite orbital angular
momentum
\eqn\lstate{
|\ell\,m\rangle = \sqrt{2\ell + 1} \int dg\ D_{m0}^{(\ell)*} (g)\ \hat R
(g) |\hat z\rangle   \  ,
}
with the spin of the light degrees of freedom to get states
\eqn\jlstate{
|F\,f; \ell\,S\rangle = \sum_{r,s} \left(\ell\,r; S\,s|F\,f\right)\
|\ell\,r\rangle |S\,s\rangle\ .
}
Using eqs.~\jstate--\jlstate\ it is straightforward to show that
\eqn\overlap{\eqalign{
\langle F'\, f'; \ell\,S|F\,f; K\,k\rangle &= \delta_{F'F}\,
\delta_{f'f}\, \delta_{KS}
\,\sqrt{{2\ell + 1\over 2F + 1}}\ \left(F\,f| \ell\,0; K\,k\right)\cr
&= \delta_{F'F}\, \delta_{f'f}\, \delta_{KS}\,
(-1)^{K+k} \left(\ell\, 0| K\, -k; F\,k\right)\ .
}}

This allows a partial wave decomposition of the eigenstates of the
potential.  Consider  for example states with $K = 1$.  Then $S =
1$ and so we can form states with orbital angular momentum $\ell = F -
1, F, F + 1$.  The other restriction is that $F \geq |k|$, so that
the $D$ matrix in eq.~\jstate\ exists.  For
definiteness, consider the case $F = 1$.  Then for $k = 0$ we have
according to eq.~\overlap\ the partial wave decomposition
\eqn\kzero{
|k = 0 \rangle = \sqrt{{1\over 3}}\ | \ell = 0\rangle -
\sqrt{{2\over 3}}\ |\ell = 2\rangle\ .
}
For $k = \pm 1$ it is convenient to form the linear combinations
\eqn\kpmdef{
| \pm
\rangle = {| k = 1\rangle\pm|k = - 1\rangle\over\sqrt{2}}\ ,
}
which decompose as
\eqn\kpm{\eqalign{
| + \rangle &= \sqrt{{2\over 3}}\ | \ell = 0 \rangle
+ \sqrt{{1\over 3}}\ | \ell
= 2 \rangle\ ,\cr
| - \rangle &= | \ell = 1\rangle\ .
}}
Table 2 gives the eigenstates of the potential and their eigenvalues
up to $F = 2$.
\midinsert
\centerline{{\bf TABLE 2}}
\nobreak\bigskip\nobreak
\begintable ~~$F$~~|~~$K = S$~~|~~$\mid k \mid$ ~
$\pm$~~|~~$\ell$~~|~~$V_\pi$
for $I = 1$~~|~~$V_\pi$ for $I = 0$~~ \crthick
0 | 0 | 0 | 0 | $-3W_0/4$ | $9W_0/4$\cr
1 | 0 | 0 | 1 | $-3W_0/4$ | $9W_0/4$\cr
2 | 0 | 0 | 2 | $-3W_0/4$ | $9W_0/4$\cr
0 | 1 | 0 | 1 | $-X_0$ | $3X_0$\cr
1 | 1 | 0 | 0,2 | $- X_0$ | $3 X_0$\cr
1 | 1 | 1+ | 0,2 | $X_1$ | $-3X_1$\cr
1 | 1 | $1-$ | 1 | $X_1$ | $-3X_1$\cr
2 | 1 | 0 | 1,3 | $-X_0$ | $3X_0$\cr
2 | 1 | 1+ | 1,3 | $X_1$ | $-3X_1$\cr
2 | 1 | $1-$ | 2 | $X_1$ | $-3X_1$\endtable
\endinsert

The ``angular momentum of the light degrees of freedom'' $F$ must be combined
with the spin of the heavy quark pair $S_Q = 0,1$ to get the total
angular momentum $J$ of the bound state. In addition, if the two heavy
quarks are of the same flavor, then only states which are completely
symmetric are allowed. This gives the additional restriction that $I +
\ell + S_Q$ is even. Since $W_0, X_0$ and $X_1$ are all positive it is
easy to identify the channels which have an attractive potential. In the
$m_Q\rightarrow\infty$ limit, the kinetic energy of the heavy mesons can
be omitted. In this case, $QQ\bar q\bar q$ bound states exist if and
only if there exists some channel in which the potential is attractive.
It is obvious from Table~2 that there exist attractive channels, so we
have another demonstration that there exist exotic states in the heavy
quark limit of QCD.

There are two $\Lambda_{{\rm QCD}}/m_Q$ corrections to the
Hamiltonian that are important for the case of heavy but finite quark masses,
such as for the $b$ or $c$ quark. The kinetic energies of the heavy mesons
\eqn\hkin{
\CH_{\rm kin} = {\phantom{a}\vec p^{\ 2}\over 2\mu} \,\, ,
}
where
\eqn\mred{
{1\over \mu} = {1\over m_{Q_{1}}} + {1\over m_{Q_{2}}} \,\, ,
}
is the reduced mass, should be included. In addition,
at order $\Lambda_{{\rm QCD}}/m_Q$ the $P^{(Q)} - P^{*(Q)}$ mass
difference $\Delta^{(Q)} = m_{P^{*(Q)}} - m_{P^{(Q)}}$ should be taken
into account.  Experimentally $\Delta^{(c)} \simeq$ 141 MeV and
$\Delta^{(b)} \simeq$ 46 MeV so for $Q=c$ and $b$ this effect is quite
significant.  We define the mass splitting so that it is zero on
$P^{(Q)}$ states and $\Delta^{(Q)}$ on $P^{*(Q)}$ states. This adds a term to
the Hamiltonian, $\hat \Delta$, that is diagonal on the
$P^{(Q)}P^{(Q)}$, $P^{(Q)}P^{*(Q)}$, $P^{*(Q)}
P^{*(Q)}$ basis. The relationship between this ``meson-type'' basis and the
$|K\,k\rangle |S_Q\,s_Q\rangle$ basis is straightforward to determine.
The meson basis is obtained by first combining the spins of the heavy
and light degrees of freedom in each $H$ particle, whereas the $K,S_Q$
basis is obtained by first combining the two heavy spins and the two
light spins. A straightforward computation of this change of basis gives:
\eqn\mesonbasis{\eqalign{
|1,1\rangle|0,0\rangle =& {1\over 2}
\Bigl\{|P^{*(Q_{1})}, 1\rangle | P^{(Q_{2})}\rangle -
|P^{(Q_{1})}\rangle | P^{*(Q_{2})}, 1\rangle\cr
&+ |P^{*(Q_{1})}, 1\rangle |P^{*(Q_{2})},
0\rangle - |P^{*(Q_{1})}, 0 \rangle
|P^{*(Q_{2})}, 1\rangle\Bigr\}\  ,\cr
\noalign{\smallskip}
|1,0\rangle |0,0\rangle =& {1\over 2} \Bigl\{|P^{*(Q_{1})},
0\rangle |P^{(Q_{2})}\rangle -
|P^{(Q_{1})}\rangle | P^{*(Q_{2})}, 0\rangle\cr
&+ | P^{*(Q_{1})}, 1\rangle |P^{*(Q_{2})},
-1\rangle -|P^{*(Q_{1})}, -1\rangle
|P^{*(Q_{2})}, 1\rangle\Bigr\}\ ,\cr
\noalign{\smallskip}
|1,-1\rangle |0,0\rangle =& {1\over 2} \Bigl\{|P^{*(Q_{1})},
-1 \rangle | P^{(Q_{2})}\rangle -
|P^{(Q_{1})}\rangle |P^{*(Q_{2})}, -1\rangle\cr
&- |P^{*(Q_{1})}, -1 \rangle | P^{*(Q_{2})}, 0\rangle +
| P^{*(Q_{1})},0\rangle |
P^{*(Q_{2})}, -1\rangle\Bigr\}\ .
}}
The mass splitting term is not diagonal in the $K,S_Q$ basis used in Table~2.
The general form of the mass splitting in this basis is complicated, and is
discussed in the Appendix. However, it is simple to compute the expectation
value of the mass splitting in a given $K$ state. For example,
\eqn\kexp{
\langle K = 1, k = \pm 1 | \hat\Delta| K = 1, k = \pm 1\rangle = {1\over 4}
\left(\Delta^{(Q_{1})} + \Delta^{(Q_{2})} + 2\Delta^{(Q_{1})} +
2\Delta^{(Q_{2})}\right)\,\, .
}

We need to obtain the energy of $B^{(*)},D^{(*)}$ ``meson molecules'' including
both the kinetic energy and the mass splitting, to determine whether they
are bound. A particularly promising entry in Table 2 is the $I = 0$,
$S_Q = 0$ state on the sixth line. This has the most attractive
potential allowed in the table, and has a small orbital angular momentum
barrier. It is also a state which is allowed when the two heavy quarks
are identical. In the next section we examine the energy of this state
in the case when both heavy quarks are $b$-quarks using a variational
calculation. This state is not an eigenstate of the Hamiltonian, since
the mass splitting and the kinetic energy are not diagonal in the $K$
basis. The mass splitting
will produce mixing to other $\ket{K\,k}\ket{S_Q\,s_Q}$ states, which will
only lower the energy further, and make the state more strongly bound.
This $F=1$ state has total angular momentum one and even parity  so it
cannot decay (strongly) to
$BB$.  As long as the expectation value of the Hamiltonian $\CH =
\CH_{\rm kin} + V + \hat\Delta$ in this state is less than
$\Delta^{(b)}$ (the mass of a widely separated $B$ and $B^*$) it is
stable with respect to the strong interactions. The state can decay
electromagnetically to $BB\gamma$ if the expectation value of the
Hamiltonian is positive. Otherwise, it can only decay by the weak
interactions.\par \indent There are $\Lambda_{QCD}/m_Q$ effects we have
neglected. For example at this order the heavy quark symmetry relation
between the $P^{*(Q)}P^{*(Q)}\pi$ and $P^{(Q)}P^{*(Q)}\pi$ couplings is
altered. This effect however is less important than those
we have included.

\noindent {\bf 4.  Variational Calculation of the Binding Energy for $Q
= b$}

In this section we consider the case where both heavy quarks are
$b$-quarks and focus on the state in line 6 of Table 2 with $I = 0$ and
$S_Q = 0$.  The energy $E$ (above $2m_B$) of this trial state is
\eqn\efnal{
E[\phi] = \int_0^\infty r^2 dr\ \phi^*\left(r\right)
\CH\phi \left(r\right)\,\, ,
}
where
\eqn\heff{
\CH = - {1\over m_B} \left[{d^2\over dr^2} + {2\over r} ~ {d\over
dr}\right] + {2\over m_Br^2} + V\left(r\right) + {3\over 2} \left(m_B^*
- m_B\right) \,\, .
}
$\phi\left(r\right)$ is a trial radial wavefunction normalized to
unity,
\eqn\phinorm{
\int_0^\infty r^2 dr\ \phi^*\left(r\right) \phi\left(r\right) = 1 \,\, .
}
The orbital angular momentum barrier follows from
eq.~\kpm. The state being considered is a linear combination of $\ell=0$
and $\ell=2$ with probabilities $2/3$ and $1/3$, so the effective value
of $L^2$ is $(0)(0+1)(2/3)+(2)(2+1)(1/3)=2$. Thus the angular momentum
barrier is the same as for an $\ell=1$ state.

At large distances the potential $V\left(r\right)$ is given by the one pion
exchange potential
\eqn\vvar{
V_\pi \left(r\right) = - 3 X_1 \left(r\right)\,\, .
}
We expect $V\left(r\right)$ to be given by eq.~\vvar\ for $r \geq r_{\rm
min} = (1/2 m_\pi)$.  However, some information on the short range part of the
potential is needed to make further progress.  In the case of nuclear
forces there is a short range repulsive core that is often attributed to
vector meson exchange \nuc.  The situation
in the heavy meson case is quite different.  Suppose the nonet of $\bar
qq$ vector mesons $V_a^{\mu b}$ is coupled to the heavy mesons via the term
\eqn\vectormeson{
{\cal L} = g_V\ \Tr\ \bar H^a H_b\, v_\mu\, V_a^{\mu b} \,\, .
}
This is the type of coupling the constituent quark model suggests is
appropriate.  The coupling in eq.~\vectormeson\ gives rise to the contribution
\eqn\rhopot{
V_{\rho,\omega} \left(\vec q\right) = {g_V^2\over \vec q^{\ 2} + m_V^2}
\left\{{1\over 2} \left(\vec I^{\ 2} - 3/2\right) + {1\over 4}\right\} \,\, .
}
to the potential, where we have taken $m_\rho =m_\omega = m_V$, and
defined $\vec I$ to be the total isospin in the two particle channel.
The first term in the braces comes
from $\rho$ exchange and the second comes from $\omega$ exchange.  Note
that $\omega$ exchange is repulsive as in the case of the nuclear
potential.  However,  while in the nucleon case the $\omega$-exchange
piece is much larger than the rho exchange piece,  for the heavy meson
potential rho exchange dominates in the $I = 0$ channel giving an {\it
attractive} potential from vector meson exchange.  There is no repulsive
hard core in the heavy meson potential. Multiple pion exchange
contributions to the heavy meson potential are also less important than
for nucleons. The pion-nucleon coupling constant is $g^2=1.56$, whereas
the heavy-meson nucleon coupling is $g^2\ltap 0.5$. This, and the fact
that the analog of the $\Delta$ resonance is not integrated out, implies that
multiple pion graphs are relevant in the heavy meson case only for much
smaller values of $r$ than in the nucleon case.

Given that vector meson exchange as modeled by eq.~\rhopot\ is attractive in
the channel we are considering, a conservative approach is to use in our
variational calculation the potential
\eqn\vapprox{
V\left(r\right) = \cases{V_\pi \left(r_{\rm min}\right)&
$r\le r_{\rm min}$,\cr\noalign{\medskip}
V_\pi \left(r\right)& $r> r_{\rm min}$,\cr}
}
corresponding to flattening out the one pion exchange potential in eq.~\vvar
\ for $r<r_{\rm min}$.  It is important to remember, however, that eq.~\vapprox
\ is a (conservative) guess and conclusions drawn from it should not
be taken too seriously.  There are physical effects that increase the
potential energy which we have neglected.  For example, $\eta$ exchange
gives the contribution
\eqn\veta{
V_\eta \left(\vec q\right) = - {2g^2\over 3f^2}\ {\left(\vec S_{\ell 1} \cdot
\vec q
\right)\left(\vec S_{\ell 2} \cdot \vec q\right)\over \vec q^{\ 2} +
m_\eta^2}\,\, ,
}
to the Fourier transform of the potential.  In the $I = 0$ channel it
has the opposite sign from pion exchange.  Its effects, however, are
quite small since in the $I = 0$ channel since it is suppressed numerically by
a factor of 1/9. In position space, there is an additional suppression factor
because the potential falls off exponentially
in a distance $m_\eta^{-1}$ rather
than $m_\pi^{-1}$.  There are also contributions from derivative vector
meson couplings to the heavy mesons.  These are less important than
eq.~\vectormeson\ at large distances but may be of comparable importance
at $r \sim 1/m_{\rho }$.  In the case of the nuclear potential their
contribution to the tensor force is thought to be very important even at
distances as large as 1~fm.

The variational calculation using the energy function eq.~\efnal\ with
Hamiltonian eq.~\heff\ and potential eq.~\vapprox\ is straightforward.
We have chosen to do the computation with $g^2$ equal to the present
experimental bound of $g^2=0.5$. A simple trial wavefunction can be
chosen of the form
\eqn\trialw{
\phi(r)= N\ e^{-a r} r^b\left(1+c r\right),
}
where $N$ is a normalization constant chosen so that $\phi$ satisfies
eq.~\phinorm.
The minimum of the energy is at $a=6.23\ m_\pi$,
$b= 2.26$ and $c= -0.16\ m_\pi$.
For these values of the parameters the wavefunction $\phi(r)$ is peaked
near $r=r_{\rm min}$. The state is bound, with a binding
energy of $8.3$~MeV relative to the $BB^*$ energy. (Recall that the
state we are considering cannot decay into $BB$.) The average radial
kinetic energy is $25.5$~MeV, the average angular momentum barrier energy is
$31.3$~MeV, and the average potential energy is $-88.2$~MeV. The mass
splitting $\Delta^{(b)}$ contributes an additional $23$~MeV of energy,
since the state is 50\% $BB^*$ and 50\% $B^*B^*$.

The binding energy is sensitive to the precise value of $g^2$. For
example, with $g^2=0.6$, it is $26.9$~MeV, whereas for $g^2=0.4$, the
state is not bound by about $8.2$~MeV. The value of the binding energy
is also sensitive to the value of the potential below $r_{\rm min}$. We
have chosen to use a flat potential below $r_{\rm min}$ as a
conservative extrapolation, and used the neutral pion mass in the
numerical computations (which gives a weaker potential).
The state could be much more strongly
bound if the potential is more negative than our estimate.  We have also
investiagated the possibility of $DB$ and $DD$ bound states. With the
approximations we have made, these states are not bound. The reduced
mass of these states is small enough that the kinetic energy overwhelms
the attraction due to the potential. However, it is possible that these
states are bound if the interaction potential is more attractive than
our estimate.

It is interesting that there  is a limit of QCD in which one can show
that there must exist states with exotic quantum numbers.  These $QQ\bar
q\bar q$ states
are very difficult to produce or detect experimentally. It is much easier
to produce a meson-antimeson bound  state. The one-pion exchange
potential for meson-antimeson bound states is the negative of the
potential for meson-meson bound states that we have studied in this
paper. The meson-antimeson spectrum can be investigated by similar
methods to those used here. There is one important difference---the
meson-antimeson sector has annihilation channels which do not exist in
the meson-meson sector, so there will be no stable bound states.
However, there might exist resonances.

We thank F.~Feruglio, N.~Isgur, S.~Koonin and H.D.~Politzer for useful
discussions. This work was supported  in part by the U.S. Dept. of Energy
under Contract no. DEAC-03-81ER40050 and Grant No. DEFG03-90ER40546 and by a
NSF Presidential Young Investigator Award PHY-8958081.

\appendix{A}{Transformation of Basis}

The transformation between the $\ket{K\,k}\ket{S_Q\,s_Q}$ basis (which gives
eigenstates of the potential) and the usual angular momentum-meson type
basis is
computed explicitly in this appendix. The states $\ket{K\,k}$ are
obtained by combining the spins of the light degrees of freedom in the
two meson, and the states $\ket{Q\,q}$ are obtained by combining the
spins of the heavy quarks in the two mesons.  (We will denote the
states $\ket{S_Q \, s_Q}$ by $\ket{Q\,q}$ from now on, to avoid
multiple subscripts in the formul\ae.) It is  useful to define the states
\eqn\ksstate{
\ket{P\,p;K\, Q} = \sum_{k,q}\left(K\, k; Q\, q|P\, p\right)\
\ket{K\,k}\ket{Q\,q}\  ,
}
where the angular momentum $P$ is the sum of $K$ and $Q$. It is
convenient to treat the spin zero meson $P^{(Q)}$ and the three possible
polarizations of the spin one meson $P^{*(Q)}$ using  a unified
notation. For this reason, let $\ket{W_1\,w_1}$ denote the first meson,
where $\ket{0\,0}$ is the spin zero $P^{(Q)}$, and $\ket{1\,w}$ denotes
the three possible polarization states of the vector $P^{*(Q)}$ meson.
The two meson state is then denoted by $\ket{W_1\,w_1}\ket{W_2\,w_2}$.
Define the state $\ket{S\,s}$ to be the state obtained by combining the
total spins of the two mesons,
\eqn\sstate{
\ket{S\,s;W_1\,W_2} = \sum_{w_1,w_2}\left(W_1\, w_1; W_2\, w_2|S\, s\right)\
\ket{W_1\,w_1}\ket{W_2\,w_2}\ .
}
The transformation formul\ae\ will involve the overlap of the two states
$\ket{P\,p;K\, Q}$ and $\ket{S\,s;W_1\,W_2}$. Now $\ket{P\,p;K\, Q}$
is obtained by first combining the light spins of the mesons into $K$
and the heavy spins of the mesons into $Q$, and the resultant into
$P$, whereas $\ket{S\,s;W_1\,W_2}$ is obtained by first combining the
light and heavy spins of the first meson into $W_1$, and of the second
meson into $W_2$, and the resultant into $S$. The overlap is therefore
\eqn\nine{\eqalign{
\langle{P\,p;K\, Q}\ket{S\,s;W_1\,W_2}&=
\sqrt{\left(2W_1+1\right)\left(2W_2+1\right)
\left(2K+1\right)\left(2Q+1\right)}\cr
\noalign{\smallskip} &
\qquad\qquad\times\left\{\matrix{1/2&1/2&W_1\cr
1/2&1/2&W_2\cr K&Q&S\cr}\right\} \delta_{PS}\delta{ps},
}}
using the definition of the 9-$j$ symbol.

The eigenstates of the potential with total angular momentum $j$ are
\eqn\jkq{
\ket{j\,m;K\, k\ Q\, q} =
\sqrt{2j+1} \int_{SU(2)} dg\ D_{m\ k+q}^{(j)*} (g)\ \hat
R (g)\left[\ | \hat z\rangle\ket{K\,k}\ket{Q\,q}\ \right].
}
These states are not identical to the ones defined in eq.~\jstate\
because we have also included the heavy quark spin $\ket{Q,q}$ along
with $\ket{K,k}$ in the definition of the state. The conventional
states are obtained by taking the spatial
wavefunctions of definite orbital angular momentum  eq.~\lstate\ and
combining them with the spin state of
the mesons given by eq.~\sstate,
\eqn\orbit{\eqalign{
&\ket{j\,m;\ell\, S\, W_1\, W_2} = \sqrt{2l+1}\ \sum_{r,s}
\left(\ell\, r; S\, s| j\, m\right)\cr
&\qquad\qquad\times\int_{SU(2)} dg\ D_{r0}^{(l)*} (g)\
\left[\ \hat R (g)| \hat z\rangle\ \right] \ket{S\,s;W_1\,W_2} \ .
}}
The transformation matrix is then
\eqn\tmatrix{\eqalign{
&\langle{j\,m;\ell\, S \,W_1\,W_2}\ket{j\,m; K \,k\, Q\, q}=
\sqrt{\left(2\ell+1\right)\left(2j+1\right)}\ \sum_{r,s}
\left(\ell\, r; S\, s| j\, m\right)\cr
&\qquad\times\int_{SU(2)} dg\ D_{m\ k+q}^{(j)*} (g)\  D_{r0}^{(l)} (g)
\ \bra{S\,s;W_1\,W_2}\hat R (g)\ket{K\,k}\ket{Q\,q}. }}
Substituting the inverse of eq~\ksstate\ into eq.~\tmatrix, inserting a
complete set of states $\ket{P'\,p';K'\, Q'}$ to the left of the
rotation operator, and using
\eqn\dmatrix{
\bra{P'\,p';K'\, Q'}\hat R (g)\ket{P\,p;K\, Q}=
D^{(P)}_{p'p}(g)\ \delta_{PP'}\ \delta_{KK'}\
\delta_{QQ'}\ ,
}
leads to the expression
\eqn\expres{\eqalign{
&\langle{j\,m;\ell\, S \,W_1
\,W_2}\ket{j\,m; K \,k\, Q \,q}=\cr
&\quad\sqrt{\left(2\ell+1\right)\left(2j+1\right)}\sum_{r,s,p,p',P}
\ \left(\ell\, r; S\,
s| j\, m\right)\left(K\, k; Q\, q|P\, p\right)\cr
&\times \int_{SU(2)} dg\ D_{m\
k+q}^{(j)*} (g)\  D_{r0}^{(l)} (g) D_{p'p}^{(P)} (g)
\ \langle{S\,s;W_1\,W_2}\ket{P\,p'; K\, Q}.
}}
Rewriting the products of $D$ matrices using the Clebsch-Gordan
decomposition,  using the orthogonality of the $D$ matrices, and using
eq.~\nine\ reduces the above expression to
\eqn\final{\eqalign{
&\langle{j\,m;\ell\, S \,W_1\,W_2}\ket{j\,m; K\, k\, Q\, q}=
\sqrt{{\left(2\ell+1\right)\over\left(2j+1\right)}}
\left\{\matrix{1/2&1/2&W_1\cr 1/2&1/2&W_2\cr
K&Q&S\cr}\right\}\cr
\noalign{\medskip}
&\qquad\times\sqrt{\left(2W_1+1\right)\left(2W_2+1\right)
\left(2K+1\right)\left(2Q+1\right)}\cr
&\qquad\times
\left(K\, k; Q\, q|S\, k+q\right)\ \left(\ell\, 0; S\, k+q|j\, k+q\right)\ .
}}

The 9-$j$ symbols are invariant under reflection about either diagonal.
In addition, the symbols are invariant under even permutations of the
rows or columns, and are multiplied by $(-1)^\Sigma$ under odd
permutations of rows or columns, where $\Sigma$ is the sum of all nine
parameters. Thus the independent
9-$j$ symbols that we need for our problem are
\eqn\ninejlist{
\eqalign{
\left\{\matrix{1/2&1/2&0\cr1/2&1/2&0\cr0&0&0\cr}\right\}
&={1\over2},\cr
\left\{\matrix{1/2&1/2&1\cr1/2&1/2&1\cr0&0&0\cr}\right\}
&={1\over2\sqrt 3},\cr
\left\{\matrix{1/2&1/2&0\cr1/2&1/2&1\cr0&1&1\cr}\right\}
&={1\over6},\cr
}
\qquad
\eqalign{\left\{\matrix{1/2&1/2&1\cr1/2&1/2&1\cr0&1&1\cr}\right\}
&={1\over\sqrt{54}},\cr
\left\{\matrix{1/2&1/2&1\cr1/2&1/2&1\cr1&1&0\cr}\right\}
&=-{1\over{18}},\cr
\left\{\matrix{1/2&1/2&1\cr1/2&1/2&1\cr1&1&1\cr}\right\}
&=0,\cr
}
}
$$
\left\{\matrix{1/2&1/2&1\cr1/2&1/2&1\cr1&1&2\cr}\right\}
={1\over{9}}.
$$
Using these values we can compute the decomposition of the various
$\ket{j\,m;K\,k\,Q\,q}$ states, \eg
\eqn\decomp{\eqalign{
\ket{1\,m; 1\,1\,0\,0}=&\frac{1}{2\sqrt{3}}\ket{1\,m; 0\,1\,1\,0}-
\frac{1}{2\sqrt{3}}\ket{1\,m; 0\,1\,0\,1}-\frac{1}{\sqrt 6}
\ket{1\,m; 0\,1\,1\,1}\cr
&-\frac{1}{2\sqrt{2}}\ket{1\,m; 1\,1\,1\,0}+
\frac{1}{2\sqrt{2}}\ket{1\,m; 1\,1\,0\,1}+
\frac{1}{2}\ket{1\,m; 1\,1\,1\,1}\cr
&+\frac{1}{\sqrt{24}}\ket{1\,m; 2\,1\,1\,0}-
\frac{1}{\sqrt{24}}\ket{1\,m; 2\,1\,0\,1}-
\frac{1}{2\sqrt{3}}\ket{1\,m; 2\,1\,1\,1}
\cr
}}
where the states on the right hand side of the equation are
$\ket{j\,m;\ell\, S\, W_1\, W_2}$ states.
The decomposition of the $\ket{j\,m}$
state with $K=1$, $k=-1$, $Q=q=0$ can be obtained from eq.~\decomp\ by
changing the sign of all the $\ell=1$ terms. Thus the state $K=1$, $k=+$
which is $1/\sqrt{2}$ times the sum of the $k=1$ and $k=-1$ states
can be decomposed as
\eqn\fdecomp{\eqalign{
\ket{1\,m; 1\,+\,0\,0}=&\frac{1}{\sqrt{6}}\ket{1\,m; 0\,1\,1\,0}-
\frac{1}{\sqrt{6}}\ket{1\,m; 0\,1\,0\,1}-\frac{1}{\sqrt 3}
\ket{1\,m; 0\,1\,1\,1}\cr
&+\frac{1}{\sqrt{12}}\ket{1\,m; 2\,1\,1\,0}-
\frac{1}{\sqrt{12}}\ket{1\,m; 2\,1\,0\,1}-
\frac{1}{\sqrt{6}}\ket{1\,m; 2\,1\,1\,1}.\cr
}}
Rewriting the $W_1$ and $W_2$ labels using the more familiar $P^{(Q)}$
and $P^{*(Q)}$ labels gives
\eqn\rewrite{\eqalign{
\ket{1\,m; 1\,+\,0\,0}=&\frac{1}{\sqrt{6}}
\ket{1\,m; 0\,1\,P^{*(Q_1)}\,P^{(Q_2)}}-
\frac{1}{\sqrt{6}}\ket{1\,m; 0\,1\,P^{(Q_1)}\,P^{*(Q_2)}}\cr
&-\frac{1}{\sqrt 3}
\ket{1\,m; 0\,1\,P^{*(Q_1)}\,P^{*(Q_2)}}
+\frac{1}{\sqrt{12}}\ket{1\,m; 2\,1\,P^{*(Q_1)}\,P^{(Q_2)}}\cr
&-\frac{1}{\sqrt{12}}\ket{1\,m; 2\,1\,P^{(Q_1)}\,P^{*(Q_2)}}-
\frac{1}{\sqrt{6}}\ket{1\,m; 2\,1\,P^{*(Q_1)}\,P^{*(Q_2)}}.\cr
}}
{}From this decomposition, it is easy to see that the state on the left
hand side is 25\% $P^{(Q_1)}P^{*(Q_2)}$, 25\% $P^{*(Q_1)}P^{(Q_2)}$
and 50\% $P^{*(Q_1)}P^{*(Q_2)}$, and is 67\% $\ell=0$ and
33\% $\ell=2$. The left hand side of
eq.~\rewrite\ was expressed in terms of angular momentum eigenstates in
eq.~\kpm, and in terms of meson states in eq.~\mesonbasis, which are
special cases of the simultaneous decomposition in eq.~\rewrite.

\listrefs
\listfigs
\insertfig{Figure  1}{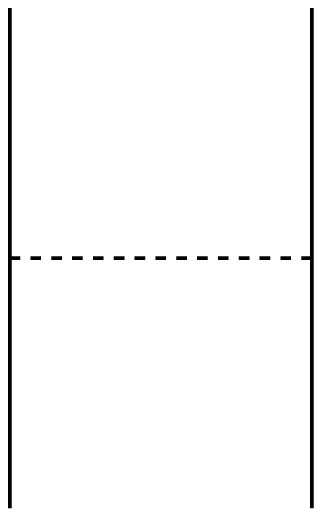}
\bye